\documentstyle[prb,aps,multicol,epsf]{revtex}

\def\fakebold#1{\relax\ifvmode\leavevmode\fi%
\ifmmode%
\setbox0=\hbox{$#1$}%
\else%
\setbox0=\hbox{#1}%
\fi%
\kern-.02em\copy0 \kern-\wd0%
\kern .04em\copy0 \kern-\wd0%
\kern-.0125em\raise.02em\box0%
}%
\newcommand{\bflambda}{\mbox{\fakebold{$\lambda$}}}

\tighten
\begin{document}
\draft
\vspace{2cm}
\title{
 Spin glass versus superconductivity}
\author{V. M. Galitski$^{1}$ and A. I. Larkin$^{1,2}$}

\address{$^1$Theoretical Physics Institute, University
of Minnesota, Minneapolis, MN 55455, USA\\
\vspace{0.5cm}
$^2$Landau Institute for Theoretical Physics,
Kossigin Str. 2, 117940, Moscow, Russia}

\maketitle
\begin{abstract}
  A superconductor with interacting paramagnetic impurities is considered.
  The impurities are coupled via the Ruderman-Kittel-Kasuya-Yoshida interaction.
  At a temperature $T_g$, the system of magnetic impurities forms a spin-glass
  state. We study the effect of the spin-spin interactions on the superconducting transition
  point at $T<T_g$. We show that  superconducting properties
  depend on the state of the spin system via spin-spin
  autocorrelation functions. With the help of the Keldysh technique,
  a general nonequilibrium Gor'kov equation is derived.
  Possible ageing effects in the superconducting transition point are
  discussed.
  The equilibrium superconducting transition point is found
  explicitly and shown to be shifted towards higher temperatures and impurity concentrations
  compared to the classical Abrikosov-Gor'kov's curve. The corresponding
  shift of the superconducting quantum critical point is quite
  small (about 10\%).
  A method of calculating spin-spin correlation function is suggested. The method
  combines the ideas of random mean-field method and virial expansion.
  We calculate analytically the first virial term for the spin-spin correlator for the quantum
  Heisenberg spin glass  with the RKKY interactions in the quasiequilibrium regime.

\end{abstract}

\begin{multicols}{2}
\section{Introduction}

The theory of superconducting alloys with paramagnetic impurities
was developed long ago by Abrikosov and Gor'kov. \cite{AG} They
have shown that the superconducting transition temperature was
suppressed by magnetic impurities. At some critical impurity
concentration, the transition temperature was suppressed down  to
zero, which gives an example of a quantum critical point.
\cite{Ram} The critical concentration can be determined from the
condition: $\tau_s T_{c0} = 2 \gamma / \pi$, where $T_{c0}$ is the
superconducting transition temperature in a sample without
impurities and $\tau_s$ is the spin-flip scattering time, which is
inverse proportional to the concentration of magnetic impurities
$n_s$.
Let us note that the effect of magnetic impurities on
superconductivity is, in many aspects, similar to the one of an
external magnetic field.\cite{HW}

In the Abrikosov-Gor'kov's model, the magnetic impurities did not
interact. Indeed, in any real system, one can not avoid having
interactions between the impurities. Friedel oscillations in the
electron density give rise to the similar oscillations in the
spin-spin coupling which is well-known as
Ruderman-Kittel-Kasuya-Yoshida (RKKY) interaction. The interaction
may change the nature of the transition both qualitatively and
quantitatively.

Interactions between impurities can significantly change the
physical picture only if the typical energy of this interaction is
of the order of temperature or higher. At a high temperature, one
can take into account only those impurity clusters in which the
typical distance between spins is small enough. The probability of
having such a cluster containing three or more spins is small.
Thus, virial expansion (or cluster expansion) coincides with the
high-temperature expansion in the problem under
discussion.\cite{LKh}  The corresponding expansion parameter is
$J_0 n_{\rm s} / T$, where $J_0$ is the amplitude of the RKKY
interaction.

Let us note that constant $J_0$ and spin-flip scattering time
$\tau_s$ both follow from the same exchange Hamiltonian.
Parametrically, they are of the same order. However, due to the
numerical smallness of the two-particle statistical weight in the
three-dimensional case, the amplitude of the RKKY interaction is
numerically much smaller than the spin-flip scattering rate: $ J_0
\tau_s n_s=1 /4 \pi^2 S \left( S + 1 \right)$, where $S$ is the
spin of a magnetic impurity.

The high-temperature corrections to the Abrikosov-Gor'kov's result
were derived in the paper of Larkin, Melnikov and Khmelnitskii.
\cite{LMK} It was shown that the superconducting transition
temperature was higher in the presence of the spin-spin
interactions compared to the non-interacting case. However, the
corresponding change was noticeable only at extremely low
temperatures due to the small numerical factor mentioned above.
Thus, in a very wide range of temperatures, the
Abrikosov-Gor'kov's theory was quantitatively correct. Let us
mention that there have been a number of experiments in which
deviations from the Abrikosov-Gor'kov's picture were observed at
low temperatures (see {\em e.g.} Ref. \onlinecite{exper}).

Since the RKKY interaction is random in sign, it introduces a
frustration into the spin system. It may result in a spin glass
transition at a low enough temperature $T_g \sim J_0 n_s$. This
makes the physical picture much more puzzled compared to the
high-temperature case. A wide-spread distribution of energy
barriers exists in the glassy phase. The typical time of classical
and quantum tunneling is comparable with the observation time in
real experiments. Thus, the state of the system depends not only
on the temperature and external magnetic field, but also on the
history. Moreover, all physical quantities slowly depend on time.
\cite{SG,SGrev}

In the present paper, we show that superconducting properties,
transition temperature $T_c$ in particular, depend on the properties
of the spin system. Thus, superconducting parameters are expected
to depend on the history and real time if the spin system is in
the ageing regime.

Although the dynamical properties of the spin glasses have been
the subject of extensive studies in the recent years, the
theoretical understanding of the effect is far from being
impressive yet. However, it is clear that, in general, the
dynamics of a glassy system consists of two parts: one is a fast
quasiequilibrium dynamics and the other is a slow dynamics or
ageing. One can expect that in the system under consideration
different superconducting parameters should acquire similar
behavior.

Studying the out-of-equilibrium dynamics in a quantum system
requires using a nonequilibrium formalism such as quantum
transport equation or Keldysh technique.\cite{Keldysh} The Keldysh
technique in superconductors was developed by Larkin and
Ovchinnikov \cite{LO} and by Feigel'man {\em et al.}
\cite{Skvortsov}. In the theory of quantum spin glasses, Keldysh
formalism was recently used by Kennett {\em et al.}\cite{KelSG} In
the theory of classical spin glasses, Keldysh technique is
replaced by so-called Doi-Peliti techniques, \cite{Doi} in which
dynamics is generated via the Langevin noise introduced in the
stochastic equation of motion. This kind of technique was used by
Ioffe {\em et al.} who considered dynamics of a classical spin
glass.\cite{Ioffe}

Our paper is structured as follows: \\
In Section II, which is necessarily quite technical, we derive
general equations on the superconducting Green functions taking
into account a possible non-equilibrium dynamics in the spin
system. The corresponding calculations are done with the help of
the Keldysh technique. We reconsider the Abrikosov-Gor'kov's
theory, taking into account inelastic exchange electron scattering
on magnetic impurities. Technically speaking, impurity lines, {\em
i.e.} spin-spin correlators, carry frequency in this case. We show
that all superconducting parameters, including the transition
temperature, should depend on the properties of the spin system
via the spin-spin autocorrelation function. In a non-equilibrium
state, the spin system is described with the help of three
correlators (retarded, advanced and Keldysh). In the
quasiequilibrium case, these correlators are connected via the
fluctuation-dissipation theorem. At the end of Sec. II, using
analytical continuation on the discrete Matsubara frequencies, we
rederive a relatively simple equation on the superconducting
transition point in the quasiequilibrium case.

In Sec. III, we address the question of calculating spin-spin
autocorrelation function in the spin-glass state. We propose a
method which combines ideas of Thouless, Anderson, and Palmer
(TAP) \cite{TAP} and virial expansion method. In the framework of
this approximation, the $N$-spin problem is solved exactly while
the other spins are replaced  by a mean value, which plays the
role of a random mean field. A distribution function for the
random mean-field is derived analytically for the case of the
 three-dimensional Heisenberg model with the RKKY interactions. Let us
note that despite the high-temperature asymptotics developed in
Ref. \onlinecite{LMK}, in the glass phase, such an expansion does
not contain any small parameter. One can expect, however, that at
large enough $N$ this expansion gives a quantitatively correct
result. At low $N$, we may expect to derive qualitatively
acceptable results and get some insight into the complicated
problem. To illustrate how the method works, we analytically
derive the spin-spin autocorrelation function in the first virial
approximation.

In Sec. IV, we derive a correction to the superconducting quantum
critical point. We show that the shift of the Abrikosov-Gor'kov's
result is quite small (the critical concentration of magnetic
impurities increases about 10\% compared to the non-interacting
case). We also discuss the back effect of the superconductivity on
the spin system. The phase diagram for the system under
consideration is given.

In Sec. V, we study a non-equilibrium or ageing dynamics in the
superconducting transition point. We propose an experiment which
should provide an explicit manifestation of the ageing dynamics in
the superconducting quantum critical point. If an external
magnetic field is switched off below $T_g$, the magnetization does
not disappear immediately. Spins magnetize the electrons and this
effect leads to a stronger suppression of superconductivity
compared to the case of arbitrarily oriented spins. The remanent
magnetization logarithmically slowly decreases and drives the
system towards superconductivity.

\section{Nonequilibrium Gor'kov equations}

\subsection{The model}

The starting point for the problem is the following
Abrikosov-Gor'kov's  Hamiltonian:
\begin{equation}
\label{H}
{\cal H}_{\rm AG} = {\cal H}_{\rm BCS} + {\cal H}_{\rm eS},
\end{equation}
where the first term is the usual BCS Hamiltonian:
\begin{equation}
\label{BCS}
{\cal H}_{\rm BCS} = \int \left\{
\psi^{\dagger}_\alpha \left(
{{\bf p}^2 \over 2m} - \varepsilon_{\rm F} \right) \psi_\alpha
-\lambda \psi^{\dagger}_\alpha \psi^{\dagger}_\beta
 \psi_\beta \psi_\alpha \right\} d^3{\bf r}
\end{equation}
and the second one is the exchange interaction:
\begin{equation}
\label{eS}
{\cal H}_{\rm eS} =
 \int \left\{
\psi^{\dagger}_\alpha \sum\limits_a u\left({\bf r} - {\bf
r}_a\right) \left( {\bf S}_a {\bf \sigma}_{\alpha \beta} \right)
\psi_\beta  \right\} d^3{\bf r}.
\end{equation}
We neglect finite-size effects and consider the following form of
the exchange potential:
\begin{equation}
\label{u0} u({\bf r}) = u_0 \delta \left( {\bf r} \right).
\end{equation}
The effective spin-spin interactions are described by the
following Hamiltonian:
\begin{equation} \label{SS} {\cal H}_{\rm
SS} = {1 \over 2} \sum\limits_{a \ne b} J\left( {\bf r}_a - {\bf
r}_b \right) {\bf S}_a {\bf S}_b.
\end{equation}
In Eq.(\ref{SS}), $J({\bf r})$ is the RKKY interaction:
\begin{equation}
\label{RKKY} J(r) = J_0 {\cos{\left(2 p_{\rm F} r \right)} \over
r^3}.
\end{equation}
The amplitude of this interaction reads:
\begin{equation}
\label{J0} J_0 = {m p_{\rm F} \over 4 \pi^3}\, u_0^2.
\end{equation}
This quantity is connected with the spin-flip scattering time as
follows:
\begin{equation}
\label{tau_s} J_0 n_s \tau_s = \left[ 4 \pi^2 S \left(S+1\right)
\right]^{-1}.
\end{equation}
Let us emphasize that the number in the right-hand side of
Eq.(\ref{tau_s}) is very small.

\subsection{Keldysh Green functions}

In order to find the superconducting transition point, we should
first derive Gor'kov's equation for the Green functions. To treat
a possible non-equilibrium dynamics, we use the Keldysh technique.
Below we recall the basic definitions and notations:

The electron Green functions are defined as follows:
\begin{equation}
\label{G>}
\hat G^{>} (1,2) =-i
\left\langle \left(
\begin{array}{cc}
\psi_{\uparrow}(1) \psi^{\dagger}_{\uparrow}(2) &
 \psi_{\uparrow}(1) \psi_{\downarrow}(2)\\
-\psi^{\dagger}_{\downarrow}(1) \psi^{\dagger}_{\uparrow}(2) &
-\psi^{\dagger}_{\downarrow}(1) \psi_{\downarrow}(2)
\end{array}
\right) \right\rangle
\end{equation}
and
\begin{equation}
\label{G<}
\hat G^{<} (1,2) =i
\left\langle \left(
\begin{array}{cc}
\psi^{\dagger}_{\uparrow}(2) \psi_{\uparrow}(1) &
 \psi_{\downarrow}(2) \psi_{\uparrow}(1)\\
-\psi^{\dagger}_{\uparrow}(2) \psi^{\dagger}_{\downarrow}(1) &
-\psi_{\downarrow}(2) \psi^{\dagger}_{\downarrow}(1)
\end{array}
\right) \right\rangle.
\end{equation}

Retarded, advanced and Keldysh Green functions are then
constructed in the following way:
\begin{equation}
\label{GR}
\hat G^{R} (1,2) = \theta\left( t_1 - t_2 \right)
\left[ \hat G^{>} (1,2) - \hat G^{<} (1,2) \right],
\end{equation}
\begin{equation}
\label{GA}
\hat G^{A} (1,2) = - \theta\left( t_2 - t_1 \right)
\left[ \hat G^{>} (1,2) - \hat G^{<} (1,2) \right],
\end{equation}
and
\begin{equation}
\label{GK} \hat G^{K} (1,2) = \hat G^{>} (1,2) + \hat G^{<} (1,2).
\end{equation}
It is convenient to introduce a compact notation for the 4x4
matrix (here we use the Larkin-Ovchinnikov representation
\cite{LO}):
\begin{equation}
\label{G4}
\check{G}(1,2) =
\left(
\begin{array}{cc}
\hat G^{R}(1,2) & \hat G^{K}(1,2) \\
0 & \hat G^{A}(1,2)
\end{array}
\right).
\end{equation}
The Green function satisfies the following matrix equation:
\begin{equation}
\label{Gsig} \left[ \check{G}_0^{-1} - \check{\Sigma} \right]
\check{G} = \check{1}.
\end{equation}
where $\check{G}_0$ is the Green function of a normal metal
without impurities and $\check{\Sigma}$ is the self-energy.

\begin{figure}
\epsfxsize=6.5cm \centerline{\epsfbox{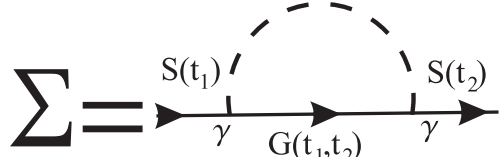}}
\caption{\label{fig:self} Electron self-energy in the leading
approximation with respect to the magnetic impurity concentration.
$\gamma$'s are the vertex matrices [see Eq.(\protect\ref{gamma})].
}
\end{figure}
In the case under consideration, we have to find the self-energy
associated with the scattering on magnetic impurities. In the
leading approximation on the impurity concentration, the
self-energy has the form shown on  Fig.~1, where $\gamma$'s are
the vertex matrices (see below). In this section, we will consider
the case of zero-field cooling, so that there is no remanent
magnetization in the spin system and no external magnetic field.
In this case, we can average Green functions out over the spin
degree of freedom. In the absence of the RKKY interaction, the
electron scattering on magnetic impurities is elastic in the Born
approximation. If we switch on the interactions between spins,
self-energy $\Sigma$ becomes non-elastic and in some sense
analogous to the self-energy due to the electron scattering off
the phonons. As a result we obtain the following expression for
the self-energy:
\begin{equation}
\label{Sig}
\check{\Sigma}_{\mu \nu} =
{1 \over 4 \tau_s S(S+1)}
\gamma^{\lambda}_{\mu \sigma}
\left( \check{\tau_z} \check{g} \check{\tau_z} \right)_{\sigma \rho}
\hat{C}_{\lambda \eta} \gamma^{\eta}_{\rho \nu},
\end{equation}
where the Greek indexes label matrix elements in the Keldysh
space. The structure of the self-energy in the Nambu-Gor'kov space
is determined by the product $\left( \check{\tau_z} \check{g}
\check{\tau_z} \right)$ with
\begin{equation}
\label{tz}
\check{\tau_z} =
\left(
\begin{array}{cc}
\hat{1} & 0\\
0 & -\hat{1}
\end{array}
\right).
\end{equation}
In Eq. (\ref{Sig}), we have introduced quasiclassical Green function:
\begin{equation}
\label{g} \check{g}(t_1,t_2) = {i \over \pi}
\int\limits_{-\infty}^{\infty} d\xi_{\bf p} \check{G}({\bf
p};t_1,t_2),\,\,\,\, \xi_{\bf p} = v_{\rm F} \left( p - p_{\rm F}
\right),
\end{equation}
and the following spin-spin correlators:
\begin{equation}
\label{C><} C^{>}(t_1,t_2) = C^{<}(t_2,t_1) = -i \left\langle {\bf
S}(t_1) {\bf S}(t_2) \right\rangle.
\end{equation}
The corresponding Keldysh matrix is defined as
\begin{equation}
\label{CKel}
\hat{C} =
\left(
\begin{array}{cc}
 C^{K} & C^{R} \\
C^{A} &  0
\end{array}
\right),
\end{equation}
with $C^R$, $C^A$, and $C^K$ defined via Eqs.(\ref{GR}--\ref{GK}).
Indeed, the spin-spin correlators are proportional to the unit
matrix in the Nambu space. The vertex matrices has the following
explicit form in the Larkin-Ovchinnikov representation: \cite{KA}
\begin{equation}
\label{gamma} \gamma^1 = \left(
\begin{array}{cc}
1 & 0\\
0 & 1
\end{array}
\right)\,\,\,\, \mbox{and} \,\,\,\, \gamma^2 = \left(
\begin{array}{cc}
0 & 1\\
1 & 0
\end{array}
\right).
\end{equation}

\subsection{Gor'kov equations}

Since we are not interested in the momentum dependence of the
Green function, it is convenient to  integrate out the
corresponding redundant degree of freedom. Using Eq.(\ref{Gsig}),
we obtain:
$$
\int d \xi_{\bf p} \left\{ \left[ \check{G}_0^{-1} -
\check{\Sigma} \right] \check{G} - \check{G} \left[
\check{G}_0^{-1} - \check{\Sigma} \right] \right\} = \check{0}.
$$
As a result, we obtain the following equation which involves
quasiclassical Green function only:\cite{LO}
\begin{eqnarray}
\nonumber
&&\left[ \check{\tau_z} {\partial \check{g} (t_1,t_2) \over \partial t_1}
+ {\partial \check{g} (t_1,t_2) \over \partial t_2} \check{\tau_z} \right]\\
\nonumber
&-& i \left[ \check{\Delta}(t_1)  \check{g} (t_1,t_2) - \check{g} (t_1,t_2) \check{\Delta}(t_2)
\right] =\\
\label{Eilen}
&-& i \left[ \check{\Sigma} \circ \check{g} - \check{g} \circ \check{\Sigma} \right] (t_1,t_2).
\end{eqnarray}
\noindent where symbol ``$\circ$'' means convolution with respect
to the time variable.

Note that each element of $\check{g}$ is a 2x2 matrix in the Nambu
space:
\begin{equation}
\label{malg}
\hat g =
\left(
\begin{array}{cc}
\alpha & \beta_1 \\
-\beta_2^* & - \alpha
\end{array}
\right),
\end{equation}
where $\alpha$ is the normal electron Green function and $\beta$
is the Gor'kov (anomalous) function. In the absence of the
external currents and/or magnetic field, $\beta_1=\beta_2$. Note
also that matrix (\ref{G4}) is  a matrix in the direct product of
independent Keldysh (time-reversal) and Nambu spaces. It is a
question of convention how to construct this matrix. One can use
either form (\ref{G4}) or another way:
\begin{equation}
\label{G42}
\check{g} =
\left(
\begin{array}{cc}
\hat \alpha & \hat \beta \\
-\hat \beta^* & - \hat \alpha
\end{array}
\right),
\end{equation}
with $\hat\alpha$ and $\hat\beta$ being matrices in the Keldysh
space. Since we are looking for the transition point, we should
first derive equation linear on the superconducting Green function
$\hat\beta$. Thus, form (\ref{G42}) is more convenient for our
purposes. Using the notations introduced above, the order
parameter $\Delta$ can be written as
\begin{equation}
\label{Delta} {\Delta}(t) = \pi \left| \lambda \right| \nu
\beta^K(t,t+0),
\end{equation}
where $\nu$ is the density of states per spin at the
Fermi-surface. In the matrix notations, the order parameter takes
on the form:
\begin{equation}
\label{DeltaK} \check{\Delta} (t)= \left(
\begin{array}{cc}
0 & \Delta(t) \\
-\Delta^*(t) & 0
\end{array}
\right).
\end{equation}
In the Keldysh space, $\hat{\Delta}$ is proportional to the unit
matrix (we neglect superconducting fluctuations).

To find the superconducting transition point we have first to
extract the equation on the Gor'kov Green function $\beta$. After
some algebra, we obtain:
\end{multicols}
\begin{eqnarray}
\nonumber &&\left[ {\partial \over \partial t_1} - {\partial \over
\partial t_2} \right] \beta_{\mu \nu} (t_1,t_2) + i \left[ \Delta (t_1)
+ \Delta (t_2) \right] \alpha_{\mu \nu}(t_1,t_2) \\ &&= -{i \over
4 \tau_s S(S+1)} \left\{ \gamma^{\lambda}_{\mu \sigma}
\gamma^{\eta}_{\rho \delta} \left[ \alpha_{\sigma \rho} C_{\lambda
\eta} \circ \beta_{\delta \nu} + \beta_{\sigma \rho} C_{\lambda
\eta} \circ \alpha_{\delta \nu} \right] (t_1,t_2) + \right.
\label{beta} \left. \left[ \alpha_{\mu \delta} \circ C_{\lambda
\eta} \beta_{\sigma \rho} + \beta_{\mu \delta} \circ C_{\lambda
\eta} \alpha_{\sigma \rho} \right] (t_1,t_2)
\gamma^{\lambda}_{\delta \sigma} \gamma^{\eta}_{\rho \nu}
\right\},
\end{eqnarray}
\begin{multicols}{2}

Although, the spin system may be out of the equilibrium state, the
electron system may be considered in equilibrium. This yields:
$$
g^K = g^R \circ f - f \circ g^A,
$$
where $f$ is the Fermi-distribution [$f(\varepsilon) =
\tanh{\left( \displaystyle \varepsilon  \displaystyle \over 2T
\right)}$ in the energy representation.]

Let us now separate slow and fast dynamics in the spin correlator
by performing Wigner transformation:
\begin{eqnarray}
\label{Wigner} \hat{C} \left(\omega, \overline{t} \right) =
\int\limits_{\infty}^{\infty} d\left(t_1-t_2\right) \hat{C}\left(
t_1-t_2,\overline{t} \right) e^{-i\omega \left( t_1 -t_2 \right)}.
\end{eqnarray}
In Eq.(\ref{Wigner}), parameter $\overline{t} =(t_1 + t_2)/ 2$
describes slow dynamics and, in some sense, is similar to the
waiting time. From
Eqs.(\ref{G4},\ref{Sig},\ref{tz},\ref{CKel},\ref{gamma},\ref{G42},\ref{DeltaK},\ref{beta}),
we obtain the following equations:
\end{multicols}
\begin{eqnarray}
\nonumber \varepsilon \beta^R(\varepsilon) - \left[\Delta(
\overline{t})\alpha^R(\varepsilon) + {1 \over 4} \Delta''(
\overline{t}) {\partial^2 \alpha^R (\varepsilon) \over \partial
\varepsilon^2}
 + \ldots  \right] = {1 \over 4 \tau_s S(S+1)}
\int\limits_{-\infty}^{+\infty}
{d \omega \over 2 \pi} \\
\times \Bigl\{ C^K(\omega) \left[ \alpha^R(\varepsilon)
\beta^R(\varepsilon - \omega) + \beta^R(\varepsilon)
\alpha^R(\varepsilon - \omega) \right]+ C^R(\omega) \left[
\alpha^R(\varepsilon) \beta^K(\varepsilon - \omega) +
\beta^R(\varepsilon) \alpha^K(\varepsilon - \omega) \right]
\Bigr\} \label{br}
\end{eqnarray}
and
\begin{eqnarray}
\nonumber \varepsilon \beta^A(\varepsilon) + \left[\Delta(
\overline{t})\alpha^A(\varepsilon)  + {1 \over 4} \Delta''(
\overline{t}) {\partial^2  \alpha^A (\varepsilon)\over
\partial \varepsilon^2}  + \ldots  \right] = {1 \over 4 \tau_s
S(S+1)} \int\limits_{-\infty}^{+\infty}
{d \omega \over 2 \pi}  \\
\times \Bigl\{ C^K(\omega) \left[ \alpha^A(\varepsilon)
\beta^A(\varepsilon - \omega) + \beta^A(\varepsilon)
\alpha^A(\varepsilon - \omega) \right]+ C^A(\omega) \left[
\alpha^A(\varepsilon) \beta^K(\varepsilon - \omega) +
\beta^A(\varepsilon) \alpha^K(\varepsilon - \omega) \right]
\Bigr\}. \label{ba}
\end{eqnarray}
\begin{multicols}{2}
The corresponding Keldysh function reads:
\begin{equation}
\label{bk}
\beta^K(\varepsilon) = \left[ \beta^R(\varepsilon) - \beta^A(\varepsilon) \right]
\tanh{\left[{\varepsilon \over 2 T}\right]}.
\end{equation}
Note, that all quantities in Eqs.(\ref{br},\ref{ba}) weakly depend
on the ``waiting time.'' Together with Eq.(\ref{Delta}),
Eqs.(\ref{br},\ref{ba}) form a closed equation system, which
contains all the information about the superconducting properties
of the physical system under consideration. The spin system is
completely  described by the three functions: $C^R$, $C^A$, and
$C^K$. These functions  are not necessarily connected with each
other in an out-of-equilibrium state and depend upon the details
of the state and history.

At the superconducting transition point, Eqs.(\ref{br},\ref{ba}) can be simplified
with the aid of the following relations:
\begin{equation}
\label{ar} \alpha^R(\varepsilon) = -  \alpha^A(\varepsilon) = 1
\end{equation}
and, thus,
\begin{equation}
\label{ak} \alpha^K(\varepsilon) = 2 \tanh{\left[{\varepsilon
\over 2 T}\right]}.
\end{equation}

\subsection{Quasiequilibrium regime}

Let us consider a regime when the dynamics of the spin glass state
has two well separated time scales: The first one is ``waiting
time'' $\overline{t}$, which is comparable with the typical time
of an experiment. At these time scales, large energy barriers
change. The second typical time  is $\hbar / T_g$. At these time
scales,  a quasiequilibrium state is reached within the large
energy barriers. Tunneling processes through the large barriers
are supposed to be very rare at such a time-scale.  In this
approximation, the waiting time can be considered as an
independent external parameter and equilibrium techniques can be
used. In this equilibrium regime the fluctuation-dissipation
theorem holds and it significantly simplifies the calculation.

In the equilibrium, Keldysh, retarded and advanced spin-spin
correlators are not independent functions anymore.  They are
connected via the fluctuation-dissipation theorem, which follows
directly from the Gibbs's distribution and in our language can be
formulated as follows:
\begin{eqnarray}
\label{FDT}
C^K(\omega) = &&\left[ C^R\left( \omega \right) -  C^A
\left( \omega \right) \right]
\coth{ \left[ { \omega \over 2 T} \right]} \\
&-&{2 \pi i} \delta \left( \omega \right) \left[ \sum\limits_{k}
e^{-{\epsilon_k \over T} }\right]^{-1} \sum\limits_{\epsilon_k =
\epsilon_j} \left| \langle k | {\bf S} | j \rangle \right|^2 e^{-
{\epsilon_k \over T}}, \nonumber
\end{eqnarray}
where $\left|\left. k \right\rangle \right.$ is a quantum state in
the spin system and $\epsilon_k$ is the corresponding energy
level. Note, that for spin correlators, the last ``static'' term
in Eq.(\ref{FDT}) does not vanish.

In the equilibrium, it is convenient to introduce Matsubara Green
functions $\beta(\varepsilon_n)$ and $\alpha(\varepsilon_n)$ which
depend on the fermion frequency $\varepsilon_n = \pi (2 n + 1) T$
and the Matsubara spin-spin correlator:
\begin{eqnarray}
\label{Mats} {\cal C} \left(\omega_m \right) =
\int\limits_{0}^{1/T} d \tau \left\langle T_{\tau} \left[{\bf
S}(\tau) {\bf S}(0) \right] \right\rangle
 e^{-i\omega_m \tau },
\end{eqnarray}
where
$$
{\bf S}(\tau) =
e^{- {\cal H} \tau} {\bf S}(0)
e^{{\cal H} \tau}
$$
and $\omega_m = 2 \pi m T$ is the bosonic Matsubara frequency.

\begin{figure}
\epsfxsize=6.5cm \centerline{\epsfbox{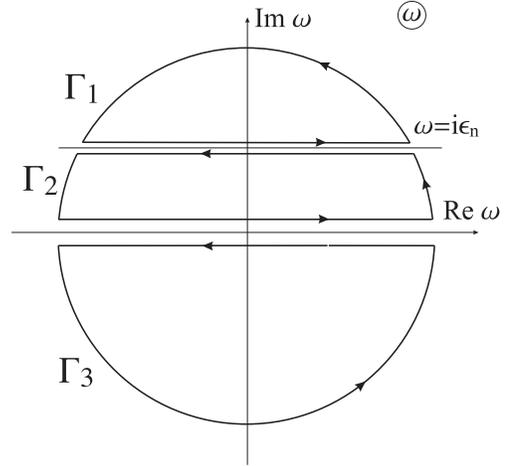}}
\caption{\label{fig:contour} Contour in the complex plane $\omega$
used to perform analytical continuation on discrete Matsubara
frequencies for $\varepsilon_n > 0$. }
\end{figure}

Using the contour shown on Fig.2, one can perform analytical
continuation on the discrete Matsubara frequencies.\cite{Mah} To
do this we first note that Keldysh spin-spin correlator consists
of two different parts: one is proportional to $\delta(\omega)$
and the other, we denote it as $\tilde{C}^K$, satisfies the
relation $\tilde{C}^K(\omega) = \left[ C^R(\omega) - C^A(\omega)
\right] \coth{\left[ \omega \over 2 T \right]}$. The former part
gives an elastic contribution to the self-energy. It is very easy
to treat this term, we just replace $\delta(\omega)$ by a delta
symbol ${1 \over T} \delta_{\omega_m\, 0}$ and the integral
(\ref{br}) by the sum over $\omega_m$. The latter part is more
complicated. As an example, let us consider the first and third
terms in the right-hand side of Eq.(\ref{br}). After analytical
continuation on discrete Matsubara frequencies $\varepsilon_n >
0$, we get for these terms:
\begin{eqnarray}
\nonumber
 \int\limits_{- \infty}^{\infty} {d \omega \over 2 \pi}
\Biggl\{ \Biggr.  \beta^R(i \varepsilon_n - \omega) \left[
 \tilde{C}^R(\omega) -  \tilde{C}^A(\omega) \right]
 \coth{\left[ \omega \over 2 T \right]} \\
 \nonumber
 +
 \tilde{C}^R(\omega) \left[
\beta^R(i \varepsilon_n - \omega) - \beta^A(i \varepsilon_n -
\omega)
\right] \tanh{\left[ i \varepsilon_n - \omega \over 2 T \right]} \Biggl. \Biggr\} \\
\nonumber =
 \oint\limits_{\Gamma_1} {d \omega \over 2 \pi}
\tilde{C}^R(\omega) \beta^A(i \varepsilon_n - \omega)
\coth{\left[
\omega \over 2 T \right]}\\
\nonumber +
 \oint\limits_{\Gamma_2} {d \omega \over 2 \pi}
\tilde{C}^R(\omega)  \beta^R (i \varepsilon_n - \omega)
\coth{\left[
\omega \over 2 T  \right]} \\
+ \oint\limits_{\Gamma_3} {d \omega \over 2 \pi}
 \tilde{C}^A(\omega) \beta^R(i \varepsilon_n - \omega)
\coth{\left[ \omega \over 2 T \right]}. \label{anconti}
\end{eqnarray}
We have used Eqs.(\ref{ar},\ref{ak}) valid at the transition point
and also the relation $ \tanh{\left[ i \varepsilon_n - \omega
\over 2 T \right]} = - \coth{\left[ \omega \over 2 T \right]}$. It
is easy to see, that in Eq.(\ref{anconti}), all $\tilde{C}$ and
$\beta$ functions  involved are analytical inside the contours of
integration. Thus, the total integral is determined by the poles
of the cotangent function only. As a result, the whole expression
(\ref{anconti}) can be rewritten as a sum over the bosonic
Matsubara frequencies:
$$
2 T \sum\limits_{\omega_m} \tilde{\cal C}(\omega_m)
\beta(\varepsilon_n - \omega_m)
$$
Similar procedure can be done with the second and forth terms in
Eq.(\ref{br}). Summarizing, we recover the quasiequilibrium
equation on the superconducting transition point.
\begin{eqnarray}
\nonumber
 \left| \varepsilon_n \right| \beta(\varepsilon_n) - \Delta
=-{T \over 2 \tau_s S \left( S + 1 \right)}
\sum\limits_{\omega_m}  {\cal C}(\omega_m) \\
\times \left\{ \beta(\varepsilon_n - \omega_m) +
\beta(\varepsilon_n)\, {\rm sgn}\, {\varepsilon_n} {\rm sgn}\,
\left( \varepsilon_n - \omega_m \right) \right\} \label{LMK}
\end{eqnarray}
Note that the equilibrium expression is the same in paramagnetic
and spin-glass phases. However, the  spin-spin autocorrelation
functions are different in these phases.

\section{Spin-spin autocorrelation function}

\subsection{Random mean-field}

To find the superconducting transition point, we have to solve
Eqs.(\ref{br},\ref{ba}) or Eq.(\ref{LMK}), which requires knowing
the spin-spin autocorrelation function(s). Calculation of this
correlator for the quantum Heisenberg spin-glass is a very hard
problem even in the quasiequilibrium case. To get some insight
into the problem under consideration, we propose a method of
calculating this correlator which introduces the concept of a
random mean field. This idea is similar to the one used by
Thouless, Anderson and Palmer. \cite{TAP}  We combine the random
mean-field method with the virial or cluster expansion [see {\em
e.g.} Ref.~\onlinecite{LMK}].

Let us consider a spin in the mean-field of the other spins
${\bf h} = \sum\limits_{a} J_{ab} {\bf S}_a$. The distribution function
for this random magnetic field can be naturally defined as follows:
\begin{equation}
\label{df1} P[{\bf h}] = \left\langle \delta \left( {\bf h} -
\sum\limits_{a} J_{ab} {\bf S}_a \right) \right\rangle_{J,{\bf S}}
\end{equation}
or, equivalently,
\begin{equation}
\label{df2} P[{\bf h}] = \int { {d^3 {\bflambda}} \over \left(
2\pi \right)^3} \left\langle \exp{\left[ - i {\bflambda} \left(
{\bf h} - \sum\limits_{a} J_{ab} {\bf S}_a \right) \right]}
\right\rangle_{J,{\bf S}},
\end{equation}
where the averaging over the random spin-spin interaction and over spin orientations
is implied. For the RKKY interaction (\ref{RKKY}), averaging means:
\begin{equation}
\label{avJ} \left\langle f(J) \right\rangle_J =: {1 \over V} \int
d^3{\bf r} \int\limits_{0}^{2 \pi} {d \phi \over 2 \pi} f \left(
{J_0 \cos{\phi} \over r^3} \right),
\end{equation}
where $V$ is the volume of the system and $f$ is an arbitrary
function.

From Eqs.(\ref{df2},\ref{avJ}), we obtain:
\begin{equation}
\label{prom} P[{\bf h}] = \int { {d^3 {\bflambda}} \over \left(
2\pi \right)^3} e^{i {\bflambda {\bf h}}} \left\langle
\prod\limits_a \left( 1 -{4 \pi \over 3} {J_0 \over V} \left|
{\bflambda}{\bf S}_a \right| \right) \right\rangle_{\bf S}.
\end{equation}
Performing averaging over the spin orientations and evaluating the
corresponding elementary integral we get the distribution function
for the random mean field:
\begin{equation}
\label{Ph} P[{\bf h}] = {1 \over \pi^2}\,  {a \over \left(a^2 +
h^2 \right)^2}, \,\,\, a= {2 \pi \over 3} J_0 n_s S,
\end{equation}
where $n_s$ is the concentration of magnetic impurities and $S$ is
their spin.

Let us note that for a different model of spin-spin interactions
the distribution function (\ref{df1}) would be different. For
example, if we start with the Gaussian distribution of spin-spin
couplings in the Sherrington-Kirpatrick model, the distribution
function of the mean field is necessarily Gaussian as well.

In the  non-equilibrium case, the simple calculation presented
above is not valid. The distribution function of the mean-field
should depend on time and on the type of the non-equilibrium
state. For example if we study a non-equilibrium dynamics due to a
slowly decaying remanent magnetization, one should modify the
definition of the distribution function as follows:
\begin{equation}
\label{df11} P[{\bf h},t] = \left\langle \delta \left( {\bf h} -
\sum\limits_{a} J_{ab} {\bf S}_a \right) \times \delta \left( {\bf
m}(t) - \sum\limits_{a} {\bf S}_a \right) \right\rangle_{J,{\bf
S}},
\end{equation}
where ${\bf m}(t)$ is the remanent magnetization.

\subsection{Virial expansion}

Now, let us consider the following set of Hamiltonians:
\begin{equation}
\label{H1}
{\cal H}_1 = {\bf S h},
\end{equation}
\begin{equation}
\label{H2}
{\cal H}_2 = J_{12} {\bf S}_1 {\bf S}_2 + {\bf S}_1 {\bf h}_1 + {\bf S}_2 {\bf h}_2,
\end{equation}
$$
\ldots\,\, ,
$$
\begin{equation}
\label{Hk}
{\cal H}^{(k)} =  {1 \over 2}{\sum\limits_{a, b}^k}' J_{ab} {\bf S}_a {\bf S}_b +
\sum\limits_{a}  {\bf S}_a {\bf h}_a,
\end{equation}
where both $J$'s and ${\bf h}$'s are random and distributed
according to Eq.(\ref{avJ}) and Eq.(\ref{Ph}), correspondingly.

In principal, one can calculate any quantity using Hamiltoninans
(\ref{H1}) (one spin), (\ref{H2}) (cluster of two spins), {\em
etc.} and than average out the quantities of interest using
Eq.(\ref{avJ}) and distribution (\ref{Ph}). This procedure
generates a series which can be called virial or cluster
expansion. It is similar to the virial expansion in the theory of
liquids and gases. Let us also note, that comparison with the
Sherrington-Kirpatrick Ising model shows that the first virial
term is equivalent to the replica symmetric solution of the model.
After a simple calculation one can obtain the equation for the
Edwards-Anderson order parameter in the replica symmetric case. It
is well known that the corresponding solution is not stable. The
usual practice is to apply the mechanism of replica symmetry
breaking.\cite{SG,SGrev} However, replica technique hardly can
lead to some explicit results especially in the dynamical
problems. Moreover, it involves a procedure of analytical
continuation on replica indexes. This procedure is usually poorly
justified, since, the behavior of functions to be analytically
continued is typically not known at large replica indexes. It is
not clear whether the virial expansion can help to solve the
underlying problems. Technically it is quite straightforward and
physically very transparent. Even though, the expansion does not
contain any small parameter at low temperatures, one can try to
calculate a quantity of interest up to some large $N$ (number of
spins in the cluster) in order to sum up the series. Even if the
series is not convergent, one can use Pad\'{e}-Borel technique to
do the summation.

If we are interested in studying a system in a non-equilibrium
regime, the Hamiltonian language described above is not
appropriate. One has to reformulate the problem using the
corresponding action on the Keldysh closed time contour and do the
virial expansion within this formalism. This idea will be
developed elsewhere.

\subsection{Matsubara spin-spin correlator. Equilibrium case.}

To give an example of how this method works, let us calculate the
spin-spin autocorrelation function in the first virial
approximation. We will consider quasiequilibrium regime only. For
the  Matsubara correlator, we obtain:
\begin{eqnarray}
\nonumber {\cal C}\left( \omega_m,h \right) = \left[ \sum\limits_k
e^{-\beta \epsilon_k} \right]^{-1}\, \sum\limits_{k_1, k_2}
\left| \langle k_1 | {\bf S}(0) | k_2 \rangle \right|^2 \times \\
\label{Gibbs} {\epsilon_{k_2} - \epsilon_{k_1} \over \omega_m^2 +
\left( \epsilon_{k_2} - \epsilon_{k_1} \right)^2} \left[ e^{-\beta
\epsilon_{k_1}} -  e^{-\beta \epsilon_{k_2}} \right].
\end{eqnarray}
For spin $S=1/2$, in the first virial approximation, we get:
\begin{equation}
\label{1:2} {\cal C}(\omega_m,h) = {1 \over 4 T}
\delta_{\omega_m,\, 0} + {h \over \omega_m^2 + h^2} \tanh{\left[ {
h \over 2T} \right]}.
\end{equation}
Note, that for an arbitrary spin value $S$, the tangent factor is replaced
by the corresponding Brillouin function.
We have to average out quantity (\ref{1:2}) with respect to the distribution
function (\ref{Ph}):
\begin{equation}
\label{mC} \overline{\cal C}(\omega_m) = \int d^3 {\bf h}\, P[{\bf
h}]\, {\cal C}(\omega_m,h).
\end{equation}
Evaluating the corresponding integral, we obtain
\begin{equation}
\label{mC2} \overline{\cal C}(\omega_m) = {1 \over 4 T}
\delta_{\omega_m,\, 0} + \overline{\delta {\cal C}}(\omega_m),
\end{equation}
where
\begin{eqnarray}
\nonumber
\overline{\delta {\cal C}}(\omega_m) = && 4 a i \left[
{\phantom{\scriptstyle .} \atop {\displaystyle {\rm res} \atop {\scriptstyle z = i a}}}
+
{\phantom{\scriptstyle .} \atop {\displaystyle{\rm res} \atop {\scriptstyle z = i \left| \omega_m \right|}}}
\right] g(z) +\\
\nonumber
&& \phantom{\scriptstyle {.}} \\
&& {2 a \over \pi} \left[ I(a,\omega_m) + 2a{\partial
I(a,\omega_m) \over \partial a} \right], \label{CT}
\end{eqnarray}
with
$$
g(z) = {z^3 \over \left( z^2 + \omega^2 \right) \left( z^2 + a^2
\right)^2} \tanh{ \left[ z \over 2 T \right] }
$$
and
$$
I(a,\omega_m) = {1 \over \omega_m^2 - a^2} {\rm Re}\, \left[
\psi\left({1 \over 2} + {i \omega_m \over 2 \pi T} \right) -
\psi\left({1 \over 2} + {i a \over 2 \pi T} \right) \right]
$$
At a low temperature $T \ll a \sim T_g$, Eq.(\ref{CT}) can be
simplified and we obtain the following zero-temperature result:
\begin{equation}
\label{dmC}
\overline{\delta {\cal C}}(\omega) =
{2 a \over \pi} {a^2 - \omega^2 + \omega^2 \ln{ \left( \omega \over a \right)^2}
\over \left( \omega^2 - a^2 \right)^2}.
\end{equation}

Let us emphasize the following property of  correlators
(\ref{mC2}) and (\ref{dmC}):
\begin{equation}
\label{a=0}
\lim\limits_{a \to 0} \overline{\cal C}(\omega_m) =
{1 \over T}\, S \left( S + 1 \right) \delta_{\omega_m \, 0},
\end{equation}
which can be verified by a straightforward calculation. This
property means that if we switch off the interactions between the
spins, the corresponding autocorrelation function turns into the
time-independent correlation function of a free spin.
%

In the second virial approximation we should calculate correlator
(\ref{Gibbs}) using eigenfunctions of Hamiltonian (\ref{H2}).
There are four quantum states possible in this case. Even at zero
temperature, finding the ground state energy requires solving
spectral problem which turns out to be an equation of the fourth
order. Moreover we should average out correlator (\ref{Gibbs})
over the random fields and coupling constant:
\begin{equation}
\label{2nd} \overline{C}(\omega_m) = \int d^3{\bf h}_1 d^3{\bf
h}_2 P[{\bf h}_1] P[{\bf h}_2] \left\langle C\left(
\omega_m;J,{\bf h}_1, {\bf h}_2 \right) \right\rangle_J
\end{equation}
This is hard to handle this analytically. Numerical work is
required in this case.

\section{Superconducting transition point}

\subsection{Quantum critical point. Equilibrium case.}
To find the superconducting transition point, we have to solve
Eqs.(\ref{br}, \ref{ba}) or Eq.(\ref{LMK}) in the quasiequilibrium
regime. Let us first consider the latter case. At a finite
temperature, Eq.(\ref{LMK}) is an infinite system of coupled
equations corresponding to different Matsubara frequencies. As
temperature goes to zero, the picture is simplified and
Eq.(\ref{LMK}) turns into an integral equation:
\begin{eqnarray}
\nonumber \left[ \left| \varepsilon \right| + {1 \over \tau_s}
\right] \beta(\varepsilon) -\Delta
 = -{1 \over 2 \tau_s S (S+1)}
\int\limits_{-\infty}^{\infty} {d \omega \over 2 \pi}\\
\nonumber \times \left[
\overline{{\cal C}}(\omega) - S \left( S +1 \right) \delta(\omega) \right]\\
\nonumber \phantom{.} \\
\times \left[ \beta(\varepsilon + \omega) + \beta(\varepsilon)
{\rm sgn}\, {\varepsilon} \,\, {\rm sgn}\, \left(\varepsilon +
\omega \right) \right]. \label{T0}
\end{eqnarray}
In the absence of spin-spin interactions the anomalous Green
function $\beta$ reads:
\begin{equation}
\label{b0} \beta^{(0)}(\varepsilon) = { \Delta \over \tau_s^{-1} +
\left| \varepsilon \right|}.
\end{equation}
Using  definition (\ref{Delta}), multiplying Eq.(\ref{T0}) on
$\beta^{(0)}(\varepsilon)$, and integrating it over $\varepsilon$,
we get:
\begin{eqnarray}
\nonumber {1 \over \pi \nu \left| \lambda \right|} -
\int\limits_{-\infty}^{\infty} {d \varepsilon \over 2 \pi}
\beta^{(0)}(\varepsilon) = -{1 \over 2 \tau_s S \left( S+1 \right)} \\
\label{pert} \times \int\limits_{-\infty}^{\infty} {d \omega \over
2 \pi}
\left[ \overline{\cal C}(\omega) - S \left( S + 1 \right) \delta(\omega) \right] \\
\times \int\limits_{-\infty}^{\infty} {d \varepsilon \over 2 \pi
\Delta} \left[ \beta(\varepsilon + \omega) + \beta(\varepsilon)
{\rm sgn}\, {\varepsilon}\,\, {\rm sgn}\, \left(\omega
+\varepsilon \right) \right] \beta^{(0)}(\varepsilon). \nonumber
\end{eqnarray}
As it will be shown below, this equation can be solved using the
perturbation theory. Due to numerical reasons, the term in the
right-hand side can be treated as a small perturbation. Thus, one
can use expression (\ref{b0}) for the Green functions in
Eq.(\ref{pert}). In this case, the integral over $\varepsilon$ in
the right-hand side of (\ref{pert}) can be easily evaluated and we
get the following expression for the integral:
\begin{eqnarray}
\nonumber
2  \tau_s I(\omega \tau_s) =2 \tau_s \Biggl\{ &\phantom{.}&{1 \over 1 + |\omega| \tau_s} \Biggr. \\
&+&
\Biggl. {1 + |\omega| \tau_s \over |\omega| \tau_s \left( 1 + |\omega| \tau_s /2 \right)}\,
\ln{ \left( 1 + |\omega| \tau_s \right)} \Biggr\}
\label{I}
\end{eqnarray}

Correlator $\overline{\cal C}$ follows from Hamiltonian
(\ref{RKKY}) and distribution function defined by relation
(\ref{avJ}). There is only one dimensional parameter upon which it
can depend: $a \sim J_0 n_s$. Thus, in the equilibrium, the
correlator is bound to have the following form:
$$
\overline{{\cal C}}(\omega) = {1 \over a} \tilde{\cal
C}\left({\omega \over a}\right)
$$
This statement is true both in the high-temperature and low-temperature regions.
Let us make the following change of variables in the integral over $\omega$ in
Eq.(\ref{pert}): $\omega\, \longrightarrow x = \omega / a$. Then, this equation
can be rewritten as follows:
\begin{eqnarray}
\nonumber
 \ln{ \left[ {\pi \over 2 \gamma} T_{c0}
\tau_s \right]}  = &\phantom{.}& {1 \over 4 \pi S \left( S + 1 \right)} \\
\label{qfteq} &\times& \int\limits_{-\infty}^{\infty} dx \left[
\tilde{\cal C}\left( x \right)- S \left( S + 1 \right) \delta(x)
\right] I(\sigma x),
\end{eqnarray}
where function $I(x)$ is defined in (\ref{I}) and $\sigma = a \tau_s$. The latter
quantity is just a number. From, Eq.(\ref{J0}) and Eq.(\ref{Ph}), we find
$$
\sigma = {1 \over 6 \pi \left(S + 1\right)} \ll 1.
$$
One can see that this number is very small. If, formally, we put $\sigma = 0$,
we will get zero in the right-hand side of Eq.(\ref{qfteq}) [see also (\ref{a=0})]
and recover the Abrikosov-Gor'kov's formula for the quantum critical
point:
$$
 T_{c0} \tau_{s0}^{\rm (AG)} = {2 \gamma \over \pi} \approx 1.13.
$$
If we keep $\sigma$ finite, the right-hand side term in
Eq.(\ref{qfteq}) will give a correction to the quantum critical
point due to the spin-spin interactions. For example, we can use
correlator (\ref{dmC}) calculated in the previous section.
Evaluating the corresponding integral numerically we get
correction to the critical concentration of magnetic impurities
concentration:
$$
{\delta n_{s0}^{(1)} \over n_{s0}^{\rm (AG)}} = 0.073.
$$
Similarly, we can obtain the shift corresponding to the second
virial approximation (\ref{2nd}). This has been done numerically.
The corresponding result is
$$
{\delta n_{s0}^{(2)} \over n_{s0}^{\rm (AG)}} = 0.085. $$ We see
that these corrections are quite small. This smallness is
connected with the parameter $T_g \tau_s$ present in the theory
and should exist in any order of the virial expansion. Thus, we
conclude that the shift of the Abrikosov-Gor'kov quantum critical
point is about $10\%$.

\subsection{Equilibrium phase diagram}

Another interesting question which is worth discussing is the
effect of the superconductivity on the effective spin-spin
interactions. This questions has been addressed long ago by
Abrikosov.\cite{Abook} Recently it was revisited by Aristov {\em
et al.}\cite{RKKY} The effective spin-spin coupling is determined
by the following expression:
\begin{equation}
\label{Jab} J({\bf r}) = 2 u_0^2 T \sum\limits_{\varepsilon_n}
\left[ \left| \alpha(\varepsilon_n, {\bf r}) \right|^2 + \left|
\beta(\varepsilon_n, {\bf r}) \right|^2 \right],
\end{equation}
where $\alpha(\varepsilon_n, {\bf r})$ and $\beta(\varepsilon_n,
{\bf r})$ are Green functions in the real space. A straightforward
calculation yields the following form of the effective RKKY
interaction below the superconducting transition point:
\begin{equation}
\label{RKKYinSC} J({\bf r}) = \left[ J_0 {\cos{ \left( 2 k_{\rm F}
r \right) } \over r^3} + {1 \over r^2}\, J_{\rm AF} \right] e^{- 2
r / \xi},
\end{equation}
where $\xi ={ v_{\rm F} / \Delta}$ and antiferromagnetic
contribution $J_{\rm AF} \propto {\Delta / \epsilon_{\rm F}}$ is
quite small.

At the transition line, $\xi = \infty$ and $\Delta = 0$, so we
recover the usual RKKY formula. However, well below the transition
the interaction becomes screened. This fact has a very transparent
physical explanation. The RKKY interactions appears due to the
fact that a spin polarizes the normal electrons around it. The
Friedel's oscillations in the electron density lead to the
Ruderman-Kittel oscillations in the spin-spin coupling. In a
superconductor, the spins are not polarized in the ground state.
Thus at the distances larger than Cooper pair size $\xi$, the
indirect spin-spin coupling must be suppressed.

\begin{figure}
\epsfxsize=8.5cm \centerline{\epsfbox{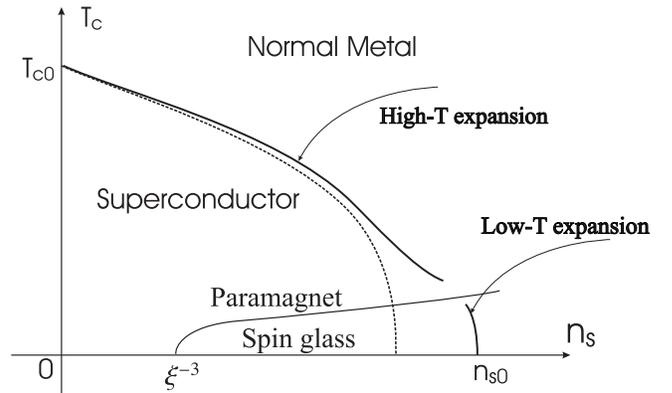}}
\caption{\label{fig:phase} Schematic phase diagram for a
superconductor with interacting magnetic impurities. The dashed
curve is the Abrikosov-Gor'kov's transition line.}
\end{figure}

In the absence of superconductivity, we expect the spin glass
state to survive even at very low concentrations of magnetic
impurities if $T=0$. In a superconductor, the spin glass state
should disappear at
$$
n_s \sim  \xi^{-3} \sim \left( {T_{c0} \over v_{\rm F}} \right)^3.
$$
We see that not only spin-spin interactions change the superconducting
phase diagram, superconductivity in turn suppresses magnetic ordering
at low temperatures and spin concentrations.

Combining these results with the ones obtained in the previous
sections, we can plot the phase diagram for the physical system
under consideration (see Fig.~3). Let us note that the
high-temperature asymptotics \cite{LMK} matches the
low-temperature one at a temperature $T^* \approx 2 J_0 n_s$,
which is approximately the temperature of the spin-glass
transition.

\section{Ageing in the superconducting quantum critical point}

Let us study the following experiment. Consider a superconductor
at a low temperature with the concentration of magnetic impurities
$n_s < n_{s0}$, where $n_{s0}$ is the equilibrium critical
concentration  found in the previous section. Now, let us switch
on a large magnetic field. This field polarizes the magnetic
impurities and electrons and destroys superconductivity. Then, we
switch off the magnetic field. The electron system equilibrate
very quickly. However, in the system of magnetic impurities, one
finds a remanent magnetization which decays very slowly with time.
This magnetization acts on the electrons as an external magnetic
field, which affects only the spin degree of freedom but not the
orbital one. In some sense the effect due to the remanent
magnetization on the electron system is equivalent to the effect
of an external magnetic field applied parallel to a
two-dimensional superconducting sample. This leads to the further
suppression of  superconductivity. Thus, even after the external
magnetic field is switched off, the  superconductivity may be
absent in such an experiment due to the ageing effects in the
system of magnetic impurities.

To get some qualitative insight into the non-equilibrium case, let
us consider a small enough remanent magnetization and neglect
corrections to the Abrikosov-Gor'kov form of the spin-spin
correlator. In this case, we obtain the following set of equations
which describes the superconducting transition and determines the
order parameter:
\begin{equation}
\label{noneq} \left[ \varepsilon - i \delta\mu \right]
\beta(\varepsilon) = \Delta \alpha(\varepsilon) - {1 \over \tau_s}
\alpha(\varepsilon) \beta(\varepsilon),
\end{equation}
where $\alpha(\varepsilon)$ and $\beta(\varepsilon)$ are
superconducting Green functions subject to the following
constraint:
\begin{equation}
\label{a+b=1} \alpha^2(\varepsilon) + \left| \beta (\varepsilon)
\right|^2 = 1
\end{equation}
and the order parameter is defined as usual:
\begin{equation}
\label{D2} \Delta = \pi \nu \left| \lambda \right| T
\sum\limits_{\varepsilon_n} \beta(\varepsilon_n).
\end{equation}
In Eq.(\ref{noneq}), we have introduced the splitting of the Fermi-surface
due to the remanent magnetization:
\begin{equation}
\label{mu} \delta \mu = u_0 m(t),
\end{equation}
where $u_0$ is the integral over the exchange potential [see
Eq.(\ref{eS})] and ${\bf m}(t) = n_s \left\langle {\bf S}
\right\rangle$ is the remanent magnetization which slowly depends
on time. Let us note, that upon $m(t)$, all superconducting
quantities in Eq.(\ref{noneq}) acquire similar slow
time-dependence.

From Eqs.(\ref{noneq}---\ref{D2}), it follows that for large
enough values of remanent magnetization $\delta \mu \sim
\tau_s^{-1}$, the superconducting transition becomes of the first
order. The order parameter jumps from the zero value up to a
finite value. Let us consider smaller values of the magnetization
and find the corresponding second-order superconducting transition
point. From Eqs.(\ref{noneq}---\ref{D2}), we get:
\begin{equation}
\label{ih+ts} 1 = \pi\nu \left| \lambda \right| T_c
\sum\limits_{\varepsilon}\, {1 \over \left| \varepsilon \right| -
i \left( \delta \mu \right)\, {\rm sgn}\, {\varepsilon} +
\tau_s^{-1} }.
\end{equation}
At zero temperature, the corresponding integral can be easily evaluated and we get:
$$
1 = \ln{ \left[ {\pi \over 2 \gamma}\, {T_{c0} \over \sqrt{
\delta\mu^2 + \tau_s^{-2}}} \right] }.
$$
Finally, we obtain the following time-dependent critical
concentration:
\begin{equation}
\label{ns(t)} n_s(t) = \sqrt{ n_{s0}^2 - c\,  m^2(t) },
\end{equation}
where constant $c$ has the form:
\begin{equation}
\label{c} c = \left[ {\pi \over S \left( S + 1 \right)}\, {u_0^2
\over m p_{\rm F}} \right]^2.
\end{equation}
The dynamics of the remanent magnetization in the spin-glass state
can be usually well-fitted via the following formula:\cite{hist}
$$
m(t) = m_0 - v \ln{t}.
$$
Where $m_0$ is of the order of the remanent magnetization just
after the magnetic field is turned off and quantity $v$ is called
magnetic viscosity. Thus,  the superconducting critical point is
slowly flowing towards its equilibrium value:
\begin{equation}
\label{nt} n_s(t) = n_{s0} - {c \over 2} \left[ m_0 - v \ln{t}
\right]^2.
\end{equation}
We see that after some time, which can be macroscopically large,
the superconductivity should appear again.

Let us note that non-equilibrium effects at the superconducting
transition point does not necessarily have to be connected with
the dynamics of remanent magnetization. Another possible
experiment  could be done as follows. Let us consider a
superconductor with the concentration of magnetic impurities such
as $n_{s0}^{\rm (AG)} < n_s < n_{s0}$, where again $n_{s0}^{\rm
(AG)}$
 is the Abrikosov-Gor'kov's critical concentration at zero temperature and $n_{s0}$ is the
real critical concentration at zero temperature with the account
of interactions. Consider large enough initial temperature so that
no superconductivity is present at the beginning. Then, let us
cool a sample down very quickly at zero field. One should expect
that superconductivity appears only some time after the sample was
cooled down due to the slow relaxation processes in the spin
subsystem.

\section{Conclusion}

We have considered a superconductor with interacting magnetic
impurities.  The central result of this paper is the dependence of
the superconducting properties in such a system upon the spin-spin
autocorrelation function. At low temperatures, one can expect to
observe ageing effects in the superconducting transition point.
The limiting manifestation of the ageing would be the observation
of the spontaneous appearance of superconductivity after some
macroscopic waiting time due to a slow change in the remanent
magnetization in the spin subsystem.

In the equilibrium, we predict that the Abrikosov-Gor'kov's
critical line shifts towards higher temperatures and impurity
concentrations due to the spin-spin interactions. Let us note that
the high-temperature expansion diverges as $T \to 0$ and,
formally, predicts very large shift of the critical point. The two
asymptotics match at a temperature $T^* \approx 2 J_0 n_s$. This
gives a good estimate for the temperature of the spin-glass
transition $T_g$. At $T = T_g$, a crossover from the
high-temperature (paramagnetic) asymptotics to the low-temperature
(spin-glass) one takes place. We have shown that the actual shift
of the superconducting transition due to the RKKY interactions is
small at any temperatures. This is connected with the small
parameter $T_g \tau_s$ which exists in the theory.

Let us also mention that our calculations are valid only for
temperatures $T \gg T_K$, where $T_K$ is the Kondo temperature. In
this case we can neglect higher order processes in the electron
scattering off the impurities. At $T \sim T_K$ the magnetic
impurities become screened, which presumably should lead to the
further enhancement of superconductivity.

Let us note that some deviations from the Abrikosov-Gor'kov's
curve have been observed experimentally.\cite{exper} The
measurements showed some increase in the critical concentration at
low temperatures. Except the effects discussed in the present
paper, this increase can also be connected with the phenomena
similar to the ones discussed in Refs. \onlinecite{isl,gal}.
Namely, optimal fluctuations in the distribution of magnetic
impurities may lead to the formation of superconducting islands
coupled via the Josephson effect. This effect may also lead to
some increase in the critical concentration. However, we do not
expect this increase to be very large due to the suppression of
the Josephson coupling by quantum fluctuations.\cite{gal} To
identify which effect is dominant at low temperatures, additional
experiments are highly desirable. In either case, we expect that
some interesting non-equilibrium phenomena (hysteresis, ageing,
{\em etc.}) should reveal themselves at low temperatures.

 \acknowledgments V. Galitski
wishes to thank Dmitry Aristov for a valuable advice. This work
was supported  by NSF grant DMR-0120702.

\end{multicols}

\end{document}